\documentclass[twocolumn,aps,showpacs,prb]{revtex4-1}

\usepackage{graphicx}
\usepackage{epstopdf}
\usepackage{amsmath}
\usepackage{multirow}

\begin{document}

\title{$\beta$-RhPb$_2$, a topological superconductor candidate}

\author{Jian-Feng Zhang$^{1}$}
\author{Peng-Jie Guo$^{1}$}
\author{Miao Gao$^{2}$}
\author{Kai Liu$^{1}$}\email{kliu@ruc.edu.cn}
\author{Zhong-Yi Lu$^{1}$}\email{zlu@ruc.edu.cn}

\affiliation{$^{1}$Department of Physics and Beijing Key Laboratory of Opto-electronic Functional Materials $\&$ Micro-nano Devices, Renmin University of China, Beijing 100872, China}

\affiliation{$^{2}$Department of Microelectronics Science and Engineering, Faculty of Science, Ningbo University, Zhejiang 315211, China}

\date{\today}

\begin{abstract}
A topological superconductor candidate $\beta$-RhPb$_2$ is predicted by using the first-principles electronic structure calculations. Our calculations show that there is a band inversion around the Fermi level at the Z point of Brillouin zone. The calculated nonzero topological invariant Z$_2$ indicates that $\beta$-RhPb$_2$ is a topological insulator defined on a curved Fermi level. The slab calculations further demonstrate that the gapless nontrivial topological surface states (TSS) are not overlapped by the bulk states and they cross the Fermi level. The phonon calculations confirm the dynamical stability of $\beta$-RhPb$_2$, while the electron-phonon coupling (EPC) calculations predict that the superconducting transition temperature ($T_c$) of $\beta$-RhPb$_2$ can reach 9.7 K. The coexistence of nontrivial topological band structure with the TSS crossing the Fermi level as well as the superconducting $T_c$ above the liquid-helium temperature suggest that the layered compound $\beta$-RhPb$_2$ is a topological superconductor, which deserves further experimental verification.
\end{abstract}


\maketitle

\section{INTRODUCTION}

The topological superconductor, characterized by a superconducting gap in bulk and Majorana zero modes at boundary, has attracted great attention recently. The Majorana zero mode is a kind of special quasiparticle that is its own antiparticle and obeys non-Abelian statistics, which possesses potential applications in topological quantum computation \cite{rmp2008,rmp2011,rpp2017}. Theoretically, as an intrinsic topological superconductor, the spinless $p+ip$ type superconductor can hold Majorana zero modes at the vortices. Nevertheless, the reported $p$-wave superconductor candidates are very scarce \cite{sro2003rmp}. On the other hand, it has been proposed that the topological superconductivity can be realized in an equivalent $p+ip$ type superconductor\cite{fu2008prl,zsc2009prl} such as the interface of a heterostructure consisting of a topological insulator (TI) and an $s$-wave conventional superconductor\cite{xqk2012sci,xqk2015prl}, where the proximity effect can induce superconductivity in the spin-helical topological surface states (TSS) \cite{fu2008prl,zsc2009prl}. Such an approach however puts forward great challenges in preparing the high-quality heterostructures and in observing the interface-related phenomena.

To avoid these difficulties in heterostructures, an alternative approach to realize topological superconductivity is to search for equivalent $p+ip$ type superconductivity in a single compound besides the spinless $p$-wave superconductors. In general, there are two strategies. The first one is to induce superconductivity in a topological insulator by charge doping as in Cu/Sr/Nb-doped Bi$_2$Se$_3$ \cite{cubise2,srbise1,nbbise1} and In-doped SnTe \cite{insnte4}. The second one is to examine the topological property of existing superconductors, such as in $\beta$-PdBi$_2$ \cite{pdbi2015nc}, PdTe$_2$ \cite{pt2017prl}, PbTaSe$_2$ \cite{pts2014prb,pts2016nc}, Fe(Te$_{0.55}$Se$_{0.45}$) \cite{fst2018sci,fst2015prb}, and A15 superconductors \cite{a15arxiv}. Here, we would like to address another strategy, i.e. we explore new compounds that possess nontrivial topological band structure as well as higher $T_c$ superconductivity.

In this work, by using the first-principles electronic structure calculations, we predict a topological superconductor candidate $\beta$-RhPb$_2$, which possesses a nontrivial topological band structure and becomes superconducting below 9.7 K. The topological surface states on the RhPb$_2$(001) surface cross the Fermi level and will give rise to a superconducting gap when the bulk becomes superconducting. Our theoretical predictions wait for experimental realization.

\section{Method}

We investigated the electronic structure, phonon spectra, and electron-phonon coupling of $\beta$-RhPb$_2$ based on the density functional theory (DFT) \cite{dft1,dft2} and density functional perturbation theory (DFPT) \cite{dfpt} calculations as implemented in the Quantum ESPRESSO (QE) package \cite{pwscf}. The interactions between electrons and nuclei were described by the norm-conserving pseudopotentials \cite{ncpp}. For the exchange-correlation functional, the generalized gradient approximation (GGA) of Perdew-Burke-Ernzerhof (PBE) \cite{PBE} type was adopted. The kinetic energy cutoff of plane-wave basis was set to be 80 Ry. The van der Waals (vdW) interactions between the $\beta$-RhPb$_2$ layers were included by using the DFT-D2 method \cite{dftd2,dftd22}. For the bulk calculations, we adopted a primitive cell with the corresponding lattice vectors ${\bf a_1}$, ${\bf a_2}$, and ${\bf a_3}$ shown in Fig. 1(a). A 12$\times$12$\times$12 {\bf k}-point mesh was used for the Brillouin zone (BZ) [Fig. 1(b)] sampling of the primitive cell. For the Fermi surface broadening, the Gaussian smearing method with a width of 0.004 Ry was employed. In the structural optimization, both lattice constants and internal atomic positions were fully relaxed until the forces on atoms were smaller than 0.0002 Ry/Bohr. To study the surface states of $\beta$-RhPb$_2$, we employed a two-dimensional (2D) supercell with a 11-layer RhPb$_2$ slab and a 20 \AA~vacuum, for which the projected 2D BZ is shown in the top part of Fig. 1(b).

The superconducting transition temperature of $\beta$-RhPb$_2$ was studied based on the electron-phonon coupling (EPC) theory as implemented in the EPW package \cite{epw}, which uses the maximally localized Wannier functions (MLWFs) \cite{mlwf} and interfaces with the QE \cite{pwscf}. We took the 4$\times$4$\times$4 for both {\bf k}-mesh and {\bf q}-mesh as the coarse grids and interpolated to the 72$\times$72$\times$72 {\bf k}-mesh and 16$\times$16$\times$16 {\bf q}-mesh dense grids, respectively. The EPC constant $\lambda$ can be calculated either by the summation of the EPC constant $\lambda_{{\bf q}\nu}$ in the full BZ for all phonon modes or by the integral of the Eliashberg spectral function\cite{Eliashberg} $\alpha^2F(\omega)$ as below,
\begin{equation}
\lambda=\sum_{{\bf q}\nu}\lambda_{{\bf q}\nu}=2\int{\frac{\alpha^2F(\omega)}{\omega}d\omega}.
\end{equation}
The Eliashberg spectral function $\alpha^2F(\omega)$ is defined as,
\begin{equation}
\alpha^2F(\omega)=\frac{1}{2{\pi}N(\varepsilon_F)}\sum_{{\bf q}\nu}\delta(\omega-\omega_{{\bf q}\nu})\frac{\gamma_{{\bf q}\nu}}{\hbar\omega_{{\bf q}\nu}},
\end{equation}
where $N(\varepsilon_F)$ is the density of states at the Fermi level $\varepsilon_F$, $\omega_{{\bf q}\nu}$ is the frequency of the $\nu$th phonon mode at the wave vector {\bf q}, and $\gamma_{{\bf q}\nu}$ is the phonon linewidth.

\begin{figure}[tb]
\includegraphics[angle=0,scale=0.29]{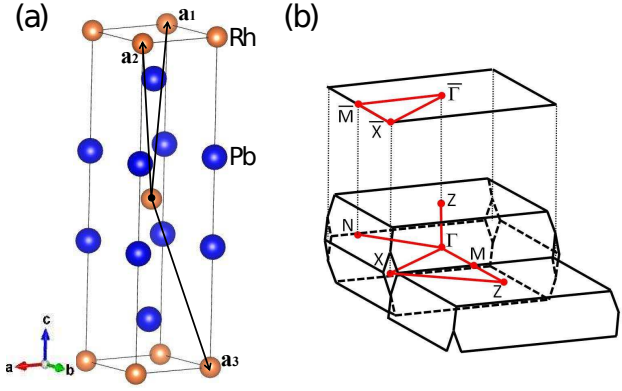}
\caption{(Color online) (a) Crystal structure of bulk $\beta$-RhPb$_2$. The blue and orange balls represent the Pb and Rh atoms, respectively. The lattice vectors ${\bf a_1}$, ${\bf a_2}$, and ${\bf a_3}$ of a primitive cell are also demonstrated. (b) The Brillouin zone (BZ) of the primitive cell and the projected two-dimensional (2D) BZ of the (001) surface of $\beta$-RhPb$_2$. The high-symmetry paths in the BZ are indicated by the red lines.}
\label{fig1}
\end{figure}

The superconducting transition temperature $T_c$ can be predicted by substituting the EPC constant $\lambda$ into the McMillan-Allen-Dynes formula \cite{mcmillan1, mcmillan2},
\begin{equation}
T_c={f_1}{f_2}\frac{\omega_{log}}{1.2}exp[\frac{-1.04(1+\lambda)}{\lambda(1-0.62\mu^*)-\mu^*}],
\end{equation}
where $\mu^*$ is the effective screened Coulomb repulsion constant, $\omega_{log}$ is the logarithmic average frequency,
\begin{equation}
\omega_{log}=exp[\frac{2}{\lambda}\int{\frac{d\omega}{\omega}\alpha^2F(\omega){log}(\omega)}],
\end{equation}
and $f_1$ and $f_2$ are the correction factors when $\lambda$ $>$ 1.3, for which the detailed calculations were presented in Ref. \onlinecite{mcmillan2}. In our calculation, $\mu^*$ was set to 0.1 in the range of the widely-used empirical values of 0.08 to 0.15 \cite{mustar1,mustar2}.

\section{Results}

\begin{figure}[tb]
\includegraphics[angle=0,scale=0.3]{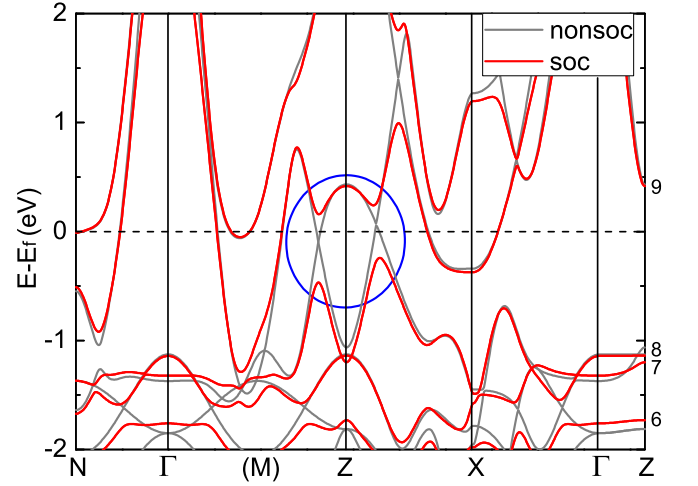}
\caption{(Color online) Band structure of bulk $\beta$-RhPb$_2$ along the high-symmetry paths in the BZ defined in Fig. 1. The gray and red lines represent the results without and with the spin-orbit coupling (SOC), respectively. The blue circle indicates the band crossing around the Z point.}
\label{fig2}
\end{figure}

Figure 1(a) shows a conventional cell of bulk $\beta$-RhPb$_2$. The $\beta$-RhPb$_2$ compound shares the same crystal structure as $\beta$-PdBi$_2$ with the space group I4/mmm instead of the CuAl$_2$-type structure with the space group I4/mcm \cite{C16}. Each Rh square lattice (labeled by orange atoms) is sandwiched by two Pb square lattices (labeled by blue atoms). The Rh atom locates at the center of eight Pb atoms, forming a RhPb$_2$ layer. Different RhPb$_2$ layers stack along $c$ direction with the AB sequence and constitute a body-centered-tetragonal (bct) structure. The optimized lattice constants are $a=b=3.25$ \AA~and $c=12.83$ \AA, and the vertical distance between Pb atoms in the same RhPb$_2$ layer is 3.38 \AA. The ${\bf a_1}$, ${\bf a_2}$, and ${\bf a_3}$ respectively indicate the lattice vectors of a primitive cell with the corresponding Brillouin zone (BZ) displayed in Fig. 1(b).

Figure 2 shows the calculated band structure of bulk $\beta$-RhPb$_2$ along the high-symmetry paths in the BZ of the primitive cell. The gray and red lines represent the results without and with the spin-orbit coupling (SOC), respectively. In the case without the SOC, there is a band crossing around the Z point near the Fermi level between the 8th and 9th bands, whose numbers are labeled at the right side of Fig. 2. These two crossing bands form a Dirac nodal ring (addressed by the blue circle), which is protected by the mirror symmetry and/or both time-reversal and space-inversion symmetries. As both Rh and Pb are heavy elements \cite{DalCorso}, it is necessary to consider the SOC effect.  Once the SOC is included, the crossing bands around the Z point gap out ($E_g$ $\geq$ 0.5 eV) and there appears a gap between the 8th and 9th bands in the whole Brillouin zone. Thus bulk $\beta$-RhPb$_2$ can be considered as an insulator defined on a curved Fermi level between the 8th and 9th bands. Given that there are both time-reversal and space-inversion symmetries in bulk $\beta$-RhPb$_2$, we can calculate its topological invariant Z$_2$ by the product of the parities of all the occupied bands at the eight time-reversal invariant momentum (TRIM) points \cite{fu2007prb}, which respectively are one $\Gamma$ point, two X points, four N points, and one Z point [Fig. 1(b)] for the BZ of bct lattice. Since the numbers of the X and N points are even, only the parities of $\Gamma$ and Z points decide the Z$_2$ topological invariant for $\beta$-RhPb$_2$. According to Table I, we can know that the Z$_2$ invariant of bulk $\beta$-RhPb$_2$ equals to 1, indicating its nontrivial topological properties.

\begin{figure}[t]
\includegraphics[angle=0,scale=0.3]{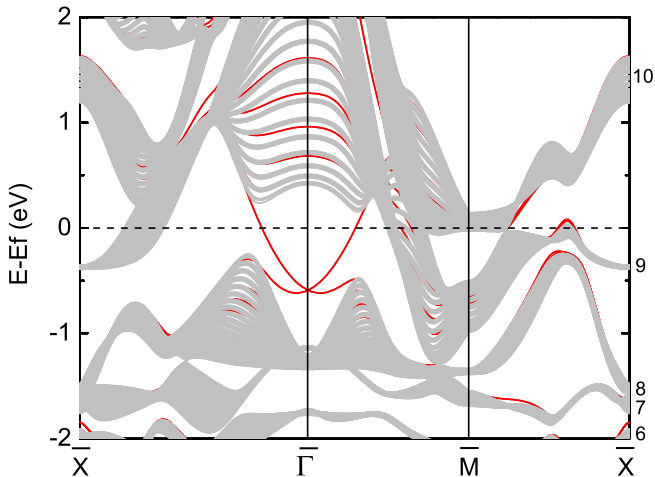}
\caption{(Color online) Band structure of the (001) surface of $\beta$-RhPb$_2$ along the high-symmetry paths in the projected 2D BZ (Fig. 1). The red and gray lines are from calculations of the 11-layer slab and the bulk of $\beta$-RhPb$_2$, respectively. The topological surface states around the $\bar{\Gamma}$ point locate within the bulk gap and cross the Fermi level.}
\label{fig3}
\end{figure}

\begin{table}[b]
\caption{The parities of all the eight occupied bands of bulk $\beta$-RhPb$_2$ below the full gap at the eight time-reversal invariant momentum (TRIM) points in the BZ of the primitive cell.}
\begin{center}
\begin{tabular*}{8cm}{@{\extracolsep{\fill}} ccccc}
\hline \hline
Parity & $\Gamma$ & 2X & 4N & Z \\
\hline
1 & + & + & - & + \\
2 & - & - & + & - \\
3 & + & + & + & + \\
4 & + & + & - & + \\
5 & + & - & + & + \\
6 & + & + & + & + \\
7 & + & - & + & - \\
8 & + & + & + & + \\
Total & - & - & + & + \\
\hline
\hline
\end{tabular*}
\end{center}
\end{table}

Besides the Z$_2$ topological invariant, the nontrivial topology can be also characterized through the surface states. Accordingly, we performed the slab calculations for the (001) surface of $\beta$-RhPb$_2$ (Fig. 3). Figures 2 and 3 show that the SOC-induced bulk band gap around the Z point is projected into the 2D BZ mainly around the $\bar{\Gamma}$ point. By comparing the band structures of slab (red lines) and bulk (gray lines) calculations, we find that the Dirac topological surface states (red lines) protected by time-reversal symmetry are not overlapped by the bulk states and they pass through the Fermi level around the $\bar{\Gamma}$ point. This will contribute to the transport properties and will also much help their observation in angle-resolved photoemission spectroscopy (ARPES) experiment.

\begin{figure}[tb]
\includegraphics[angle=0,scale=0.3]{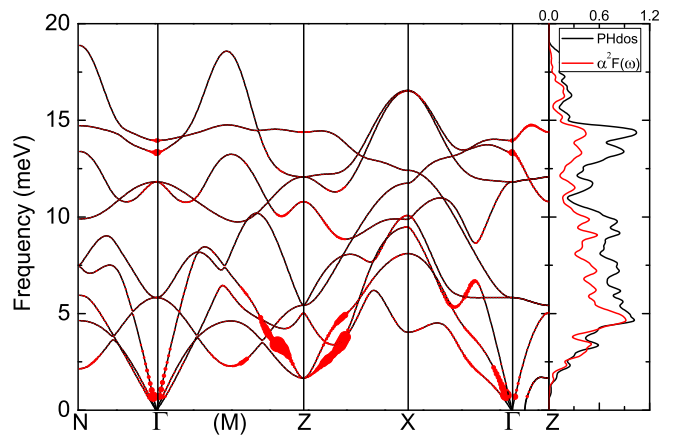}
\caption{(Color online) Phonon dispersion for the primitive cell of bulk $\beta$-RhPb$_2$ calculated with the SOC. The sizes of red dots on black lines correspond to the strengths of electron-phonon coupling (EPC) $\lambda_{{\bf q}v}$. The right panel shows the phonon density of states (black line) and the Eliashberg spectral function $\alpha^2F(\omega)$ (red line).}
\label{fig4}
\end{figure}

Considering that the $\beta$-RhPb$_2$ compound with the space group I4/mmm (Fig. 1) is not synthesized in experiment, we performed the phonon calculations to verify its dynamical stability. As shown in the phonon dispersion (left panel of Fig. 4), there is no imaginary frequency in the whole BZ except for only a tiny imaginary frequency (less than 1 meV) around the $\Gamma$ point along the $\Gamma$-Z direction. The emergence of the tiny imaginary frequency around the $\Gamma$ point can be interpreted by the difficulty of accurate interpolation in the long-range and low-frequency region. It is usually acceptable in layered materials and will not have influence on the following electron-phonon coupling (EPC) calculations.

We further performed the EPC calculations with the SOC to study the superconducting properties of $\beta$-RhPb$_2$. The calculated total EPC constant $\lambda$ of $\beta$-RhPb$_2$ is 1.76, indicating a strong EPC. In Fig. 4, the contributions to the EPC from the different phonon modes $\lambda_{{\bf q}\nu}$ are presented to be proportional to the sizes of red dots on the phonon dispersion lines (the left part). Obviously, the largest contribution comes from the acoustic branches around the Z point, which also result in several high peaks (around 3-5 meV) in the Eliashberg spectral function $\alpha^2F(\omega)$ (the right part of Fig. 4). Interestingly, the position of the {\bf q}-vector around the Z point, whose acoustic branch contributes most to $\lambda$ (Fig. 4), is very similar to the position of the {\bf k}-vector for the nodal-ring in the electronic band structure (Fig. 2). Through the Eliashberg spectral function $\alpha^2F(\omega)$ in Fig. 4 and Eq. 4, we obtain the logarithmic average of the phonon frequency $\omega_{log}$ = 7.86 meV. Furthermore, based on the McMillan-Allen-Dynes formula (Eq. 3), we predict the superconducting $T_c$ of $\beta$-RhPb$_2$ to be 9.7 K, where the correction factors $f_1$ and $f_2$ are calculated to be 1.11 and 1.05, respectively. In the case without the SOC, the total EPC constant $\lambda$ is 1.26 and the predicted $T_c$ is 6.1 K. The reduced EPC constant $\lambda$ in the absence of the SOC may originate from the giant influence of the SOC effect on the band structure around the Fermi level (Fig. 2).

We have also examined many other isostructural compounds with equal or one less valence electron, including RhGe$_2$, RhSn$_2$, IrGe$_2$, IrSn$_2$, IrPb$_2$, AgGa$_2$, AgIn$_2$, AgTl$_2$, AuGa$_2$, AuIn$_2$, AuTl$_2$, RuGe$_2$, RuSn$_2$, RuPb$_2$, OsGe$_2$, OsSn$_2$, OsPb$_2$, PtGa$_2$, PtIn$_2$, PtTl$_2$, PdGa$_2$, PdIn$_2$, and PdTl$_2$. Unfortunately, these compounds are either dynamically unstable or hold predicted superconducting $T_c$ below 3 K.

\section{Discussion and Summary}

Both the nontrivial topological band structure and the superconducting transition temperature 9.7 K are predicted for the layered-structure compound $\beta$-RhPb$_2$ by using the first-principles electronic structure calculations. Due to the proximity effect, the bulk superconductivity will further induce superconductivity in the Dirac surface states.  Such a superconductor is a topological superconductor, which is expected to hold Majorana zero modes at the vortices.

Compared with the reported potential topological superconductors, $\beta$-RhPb$_2$ has many advantages. First, as a topological superconductor, $\beta$-RhPb$_2$ avoids the difficulties in sample synthesis as those in doped topological insulators and heterostructures, and eliminates the effects of disorder or distortion. Second, the superconducting transition temperature $T_c$ of $\beta$-RhPb$_2$ (9.7 K) is above the liquid-helium temperature and higher than those of most other doped or intrinsic topological superconductor candidates: Cu/Sr/Nb-doped Bi$_2$Se$_3$ (below 4 K) \cite{cubise1,srbise1,nbbise1}, In-doped SnTe (below 4.6 K) \cite{insnte1,insnte2}, PdTe$_2$ (below 2 K) \cite{pt2017prl}, PbTaSe$_2$ (3.8 K)\cite{pts2016nc}, $\beta$-PdBi$_2$ (5.3 K) \cite{pdbi2015nc}, and Ta$_3$Sb (0.7 K) \cite{a15arxiv}, even though the $T_c$ of $\beta$-RhPb$_2$ is below that of FeSe$_{0.45}$Te$_{0.55}$ (14.5 K) \cite{fst2018sci}. Third, the giant SOC effect in $\beta$-RhPb$_2$ due to the heavy elements of component atoms induces a 0.5-eV band gap around the Z point of the 3D BZ (projected into the 2D BZ mainly around the $\bar{\Gamma}$ point) so that the inside topological surface states can be distinguished easily, which is conducive to the observation in ARPES experiment.

In summary, a topological superconductor candidate $\beta$-RhPb$_2$ has been predicted by using first-principles electronic structure calculations. The nonzero topological invariant Z$_2$ of $\beta$-RhPb$_2$ indicates its nontrivial topological property. Based on the EPC calculations, the superconducting transition temperature $T_c$ of $\beta$-RhPb$_2$ is predicted to be 9.7 K. Since the nontrivial topological surface states cross the Fermi level without overlapping with the bulk states, the superconductivity in bulk $\beta$-RhPb$_2$ will further induce superconductivity in the TSS via the proximity effect. The $\beta$-RhPb$_2$ may thus be an appropriate platform for exploring the exotic properties of topological superconductor and Majorana zero mode in experiment.

\begin{acknowledgments}

We thank Z. X. Liu and X. Liu for helpful discussions. This work was supported by the National Key R\&D Program of China (Grant No. 2017YFA0302903) and the National Natural Science Foundation of China (Grants No. 11774422 and No. 11774424). M.G. was supported by Zhejiang Provincial Natural Science Foundation of China (Grant No. LY17A040005). Computational resources were provided by the Physical Laboratory of High Performance Computing at Renmin University of China.

\end{acknowledgments}

\end{document}